\documentclass[pra,twocolumn,superscriptaddress]{revtex4}
\usepackage{amsmath}
\usepackage{amsbsy}

\usepackage{graphicx}
\usepackage{dsfont}
\begin{document}

\title{Device-Independent Randomness Generation in the Presence of Weak Cross-Talk}

\author{J. Silman}

\author{S. Pironio}
\author{S. Massar}
\affiliation{Laboratoire d'Information Quantique, Universit{\' e}
Libre de Bruxelles (ULB), 1050 Bruxelles,
Belgium}

\begin{abstract}
\textbf{Device-independent protocols use nonlocality to certify that they are performing properly. This is achieved via Bell experiments on entangled quantum systems, which are kept isolated from one another during the measurements. However, with present-day technology, perfect isolation comes at the price of experimental complexity and extremely low data rates. Here we argue that for device-independent randomness generation -- and other device-independent
protocols where the devices are in the same lab -- we can slightly relax the requirement of perfect
isolation, and
still retain most of the advantages of the device-independent approach, by allowing a little cross-talk between the devices.
This opens up the possibility of using existent experimental systems
with high data rates, such as Josephson
phase qubits on the same chip, thereby bringing device-independent randomness
generation much closer to practical application.}
\end{abstract}

\maketitle

\emph{Introduction} --
The great advantage of device-independent (DI) protocols is their reliance on a small set of tests, which are nevertheless sufficient to certify that they are performing properly. This is achieved by carrying out nonlocality tests on entangled quantum systems. In particular, no assumptions
are made regarding the inner workings of the devices, such as the
Hilbert space dimension of the underlying quantum systems, etc.
\cite{Mayers,Acin}. Each device is treated
as a `black box' with knobs and registers for selecting and displaying
(classical) inputs and outputs.  
Applications include quantum key-distribution
\cite{Barrett,Acin,McKague,Masanes,Reichardt,Vazirani}, coin flipping
\cite{Silman}, state tomography \cite{Bardyn,McKague 3,Reichardt}, genuine multi-partite entanglement detection \cite{Bancal}, self-testing of
quantum computers \cite{Magniez,McKague 2}, as well as DI randomness
generation (RG) \cite{Colbeck,Pironio,Pironio 2,Fehr,Vazirani 2}. 

It is often remarked that DI cryptographic protocols
remain secure even if the devices have been provided, or sabotaged,
by an adversary. This scenario, while conceptually fascinating, is
of little (if any) practical relevance. This is because (i) there are many types
of attacks available to a malicious provider -- the majority being classical -- that eliminating them all
is an enormous task; (ii) in any case we assume
the existence of honest providers of e.g. the source of randomness, the jamming technology to prevent information leakage
from the labs, or the classical devices used to process the data.  A scenario where one can trust the latter, but
an honest provider for the quantum devices cannot be found, is highly implausible. 

The actual
advantage of DI protocols is that they allow us
to monitor the performance of the devices irrespectively of noise,
imperfections, lack of knowledge regarding their inner workings, or
limited control over them. Indeed, even if the devices were obtained
from a trusted provider and thoroughly inspected, many things can
still unintentionally go wrong (as demonstrated by the attacks on
commercial quantum key-distribution systems \cite{Zhao,Xu,Lydersen},
which exploited unintentional design flaws). 

This problem is particularly acute in the case of DI RG, as it is very difficult even for honest parties to maufacture
reliable randomness generators (whether classical or quantum) and monitor them for malfunction.
The generation of randomness in a DI
manner solves many of the shortcomings of customary RG protocols, since, as mentioned above, the degree of violation of a Bell inequality provides
an accurate estimate of the amount of randomness generated irrespectively
of experimental imperfections and lack of control.  DI RG has so far been proven secure
against adversaries with classical side-information about the devices
 (which is the relevant case when the provider
is trusted) for arbitrary Bell inequalities and degrees of violation \cite{Pironio 2,Fehr},
and against adversaries with quantum side-information in the case
of very high violation of the CHSH inequality \cite{Vazirani 2}.

Unfortunately, DI RG is experimentally highly
challenging. It requires a Bell experiment with the detection loophole
 closed and with the quantum systems isolated from one another.

A proof of principle experiment was reported in \cite{Pironio} using
two ions in separate vacuum traps, but this system operates at
an extremely low rate ($\sim 1\,\mathrm{mHz}$), precluding any practical
application. Nevertheless, there exist today experiments involving,
for example, two Josephson phase qubits on the same
chip coupled by a radio frequency resonator \cite{Ansmann}, or two ions in the same trap coupled via their vibrational
modes \cite{Rowe,Monz}, which
allow for Bell violating experiments (with the detection loophole closed)
at much higher data rates ($\gtrsim 1\,\mathrm{kHz}$).

In these experiments the quantum systems are very close to one another.
This proximity provides the non-negligible coupling required for high entanglement
generation rates. Adapting DI RG to these types
of experiments would bring it much closer to real-life
application. The problem is that precisely because the systems are
close to one another and non-negligibly coupled, they can no longer
be considered as completely isolated (see \cite{Martinis} and \cite{Haffner}
 for a discussion of the couplings involved). The aim
of the present work is to show how to take this coupling into account
by relaxing slightly the assumptions behind the DI
approach, while keeping as much as possible all of its advantages.

We begin by showing how to derive bounds on the RG rate
in a DI setting given a known amount of cross-talk (CT).
Next, we present methods for estimating the amount of CT present in an
experiment. Our approach is then illustrated on 
Josephson phase qubits, showing that efficient DI RG is possible using already established technology. We  start first, however, by recalling briefly the essential ingredients of DI RG relevant to our analysis. We refer to \cite{Pironio,Pironio 2, Fehr} for a more detailed presentation.

\emph{Bell inequalities and device-independent randomness generation} --
A Bell experiment is characterized by the probabilities $\mathcal{P}=\left\{ P_{ab\mid xy}\right\} $
of obtaining the outcomes (or outputs) $a$ and $b$ given the measurement
settings (or inputs) $x$ and $y$. A Bell expression $\mathcal{I}\left(\mathcal{P}\right)=\sum_{abxy} c_{abxy} P_{ab|xy}$
is a linear function of these probabilities. For instance, the CHSH
inequality has the form $\mathcal{I}\left(\mathcal{P}\right)=\sum_{a,\, b,\, x,\, y\in\left\{ 0,\,1\right\} }\left(-1\right)^{a\oplus b\oplus xy}P_{ab\mid xy} \leq2$. To any Bell expression, one can associate a bound on the randomness of the outputs given the inputs $x$ and $y$ through a function $P_{xy}^{*}\left(I\right)$ such that $\max_{a,\, b}P_{ab\mid xy}\leq P_{xy}^{*}\left(I\right)$ holds for any $\mathcal{P}$ for which $\mathcal{I}\left(\mathcal{P}\right)=I$ \cite{Pironio,Pironio 2, Fehr}. The function $P_{xy}^{*}\left(I\right)$
should be monotonically decreasing and concave in $I$ (if not we
can take its concave hull). Higher values of $P_{xy}^{*}(I)$ imply less randomness, in particular when $\min_{x,\, y}P_{xy}^{*}(I)=1$
the system is fully deterministic. 

 Given knowledge of such a function and the degree of Bell violation $I$ observed in an experiment where the devices are used $n$ times in succession, one can infer a lower bound on the min-entropy of the measurement outcomes. By applying a randomness extractor
to the resulting string of outcomes, one then obtains a new private string
of random numbers of length $\simeq -n\log_2 P^*_{xy}(I)$ which is arbitrarily close (up to a security
parameter) to the uniform distribution. Depending on the assumptions made regarding the devices and the adversary, such a protocol may also require an initial random seed (in which case one talks about DI randomness expansion) that may be polynomially \cite{Pironio} or exponentially \cite{Pironio 2,Fehr} smaller than the output string. 

\emph{Device-independent randomness generation with weak cross-talk} -- 
In the security analysis of DI RG protocols the assumption that the two Bell violating devices are isolated from one another only appears in the derivation of the bound $P_{xy}^{*}\left(I\right) \geq \max_{a,\, b}P_{ab\mid xy}$. If we introduce a similar bound $P_{xy}^{*}\left(I,\chi\right)$ that is valid in the presence of a given amount of CT $\chi$ (defined below), then the rest of the reasoning of \cite{Pironio,Pironio 2,Fehr} will apply without modification. 

To define such a CT-dependent bound, we write the probabilities observed in a Bell experiment as $P_{ab|xy}=\mathrm{Tr}\left(\rho\Pi_{ab|xy}\right)$, where $\rho\in \mathcal{H}_{A}\otimes \mathcal{H}_B$ and $\{ \Pi_{ab|xy}\} $
is a POVM on $\mathcal{H}_{A}\otimes\mathcal{H}_{B}$ (i.e. $\Pi_{ab|xy}\succeq0$
and $\sum_{ab}\Pi_{ab|xy}=\mathds{1}$). The novelty with respect to the standard mathematical description of Bell experiments is in allowing the measurement $\Pi_{ab|xy}$ to act collectively on the two systems. We will say that such a collective measurement requires
no more than $\chi$ amount of CT if there exists a product POVM
$\{ \Pi_{a|x}\otimes\Pi_{b|y}\} $ satisfying \begin{equation}
-\chi \mathds{1}\preceq\Pi_{ab|xy}-\Pi_{a|x}\otimes\Pi_{b|y}\preceq\chi \mathds{1}\label{prox}\end{equation}
for all combinations of $a$ and $b$. This condition restricts how far each collective POVM may be from a product of two independent POVMs. In particular, when $\chi=0$
the $\Pi_{ab|xy}$ can be expressed as products, while when $\chi=1$
they are unconstrained.

Consider now a fixed value of $\chi$ and a Bell violation $I$. Then the
solution of the following program provides 
the minimal amount of randomness $P^*_{xy}(I,\chi)$ compatible with $I$ and $\chi$:
\begin{gather}
P_{xy}^{*}(I,\chi)  =  \max_{a,\, b}\max_{Q}\quad P_{ab\mid xy} \label{bound 1}\\
\mathrm{s.t.} \quad P_{ab\mid xy}=\mathrm{Tr}\left(\rho\Pi_{ab|xy}\right),\quad \mathcal{I}\left(\mathcal{P}\right)=I, \nonumber\\
  -\chi \mathds{1}\preceq\Pi_{ab|xy}-\Pi_{a|x}\otimes\Pi_{b|y}\preceq\chi \mathds{1}\nonumber\,,\end{gather}
where the optimization runs over the set $Q=\{\rho,\,\{\Pi_{a|x}\},\,\{\Pi_{b|y}\},\,\{\Pi_{ab|xy}\},\,\mathcal{H}_{A},\,\mathcal{H}_{B}\}$ specifying the state, measurements, and the Hilbert spaces. This formulation is therefore DI in spirit, since the 
bound is formulated without fixing the dimension  of the Hilbert spaces, nor how the measurements are implemented, etc.

\begin{figure}[t!]
\center{\includegraphics[scale=0.2]{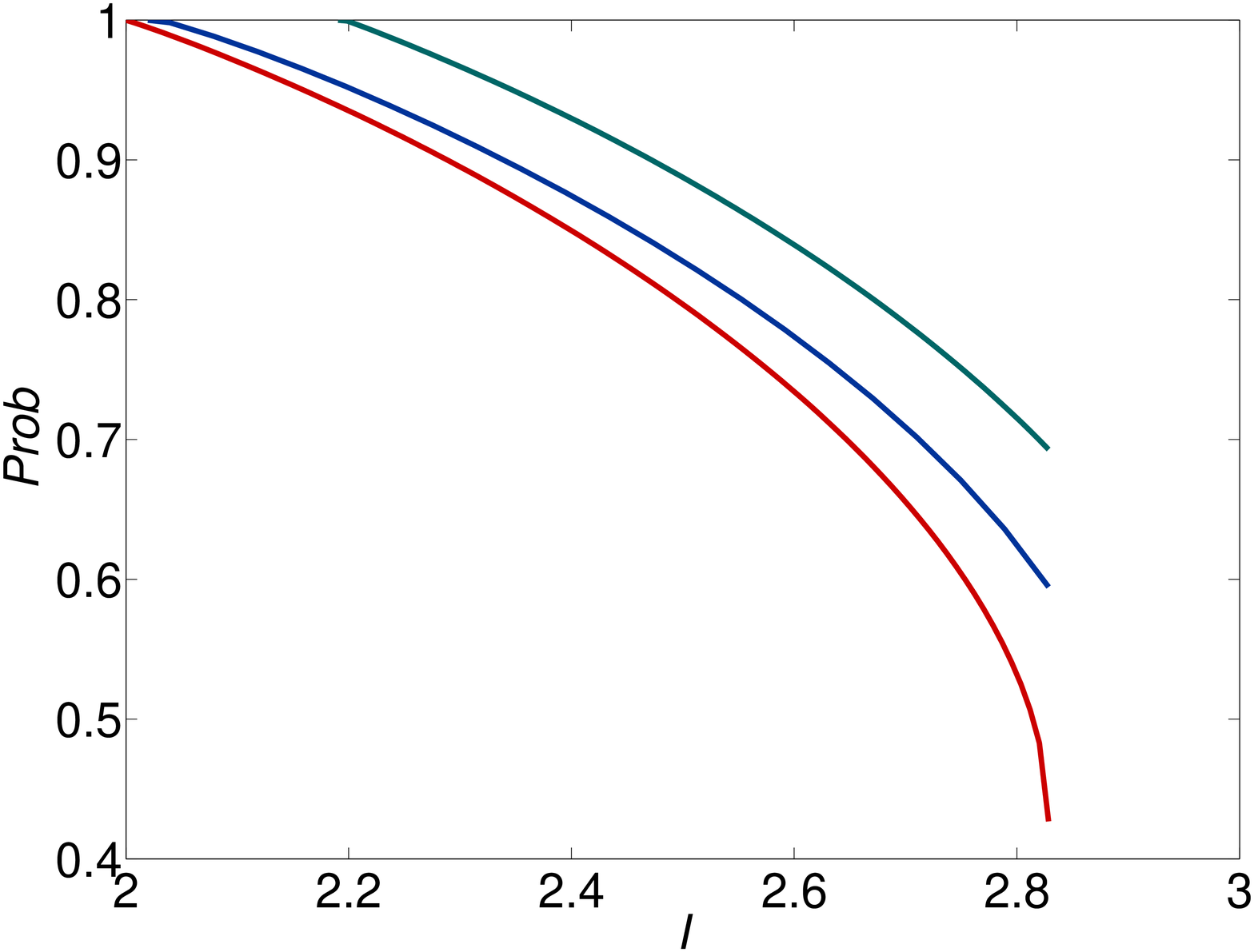}}

\caption{DI upper bounds on $P^{*}_{00}$. The middle (top) curve gives an SDP based upper bound obtained from Eq. (\ref{bound 1}) (Eq. (\ref{bound 2})), as a function of the CHSH violation $I$, given $\chi=0.01$. The bottom curve  bounds $P^{*}_{00}$ when $\chi=0$. }
\end{figure}

Upper bounds on the optimization problem Eq. (\ref{bound 1}) 
can be obtained using the 
techniques of \cite{Navascues,Pironio 3}, which relax the problem to a hierarchy
of semi-definite programs (SDPs). In particular, the resulting series
of bounds is guaranteed to converge to the true solution.
Nevertheless,
depending on the problem, even the lowest
order relaxation may be computationally intractable. We may then obtain a weaker bound  in terms of $P_{xy}^{*}\left( I,\,0\right)$ -- the solution in the absence of CT. Let 
 $\rho'$, $\{ \Pi_{a | x}' \}$, $\{ \Pi_{b| y}' \}$, and $\{ \Pi_{ab | xy}' \}$ be the state and POVMs corresponding to the solution of Eq. (\ref{bound 1}), and let $P'_{ab\mid xy}=\mathrm{Tr}(\rho' \Pi_{a | x}'  \otimes \Pi_{b | y}' )$ and $\mathcal{P}'=\{P'_{ab|xy}\}$. From the last constraint in Eq. (\ref{bound 1}) we have that $ | P_{ab\mid xy}-P'_{ab \mid xy} | \leq \chi$,
and so $\mathcal{I}(\mathcal{P}') \geq \mathcal{I}(\mathcal{P})- \gamma\chi$ where $\gamma=\sum_{a,\,b,\,x,\,y}\left|c_{abxy}\right|$ (in the case of the CHSH inequality for instance $\gamma=16$). Taken together, the last two inequalities imply that \begin{equation}
P_{xy}^{*}\left(I,\,\chi\right)\leq P_{xy}^{*}\left(I-\gamma\chi,\,0\right)+\chi\,.\label{bound 2}\end{equation}
Fig. 1 displays upper bounds on $P_{00}^{*}$ obtained from Eq. (\ref{bound 1}) and Eq. (\ref{bound 2}) in the case of the CHSH inequality.

Finally, we note that the last constraint in Eq. (\ref{bound 1}) implies that the signaling -- the extent to which the output of one device depends on the input of the other -- is constrained.  Specifically, if to each input $x$ and each input $y$ correspond $N$ outputs, $|P_{a|xy}-P_{a|xy'}|\leq 2N\chi$ for all $a,\,x,\,y,\,y'$, etc. (in the case of zero signaling, one has $P_{a|xy}=P_{a|xy'}$). This allows us to derive a simpler  bound on $P^{*}_{xy}$, depending solely on the amount of signaling present, in contrast to the bounds Eqs. (\ref{bound 1}) and (\ref{bound 2}), which rely
on the full structure of quantum mechanics. To this end we define the maximal amount of signaling allowed as 
\begin{equation}\label{sig}
\delta = \max\Bigl\{\max_{a,\,x,\,y,\,y'} |P_{a|xy}-P_{a|xy'}|,\,\max_{b,\,y,\,x,\,x'} |P_{b|xy}-P_{b|x'y}|\Bigr\}\,.
\end{equation}
When $\delta=0$, $\mathcal{P}$ resides within the no-signaling polytope \cite{Barrett 2},
while when $\delta>0$ $\mathcal{P}$ resides within a larger, higher-dimensional polytope. The bound can be obtained by solving the linear program $P^*_{xy}(I,\,\delta)=\max_{ab}P_{ab|xy}$, given that $I(\mathcal{P})=I$, $|P_{a|xy}-P_{a|xy'}|\leq \delta$, and $|P_{b|xy}-P_{b|x'y}|\leq \delta$. In the case of the CHSH inequality, one can show that (see Appendix A)
\begin{equation}
P^{*}_{xy}(I,\delta)\leq\frac{3}{2}-\frac{1}{4}I+2\delta\,.\label{bound 3}\end{equation} 
This bound applies to any post-quantum theory which restricts the
amount of signaling (as well as to quantum mechanics).

\emph{Estimating the amount of cross-talk} -- 
We have just seen how the introduction of a new security parameter $\chi$, quantifying the amount of CT between the devices, allows us to extend the scope of DI RG to settings with a limited amount of CT. To apply this approach, we therefore need a reliable prior estimate of $\chi$, and means of guaranteeing or verifying that the CT will not exceed this estimate during latter operations of the devices. This obviously requires some modeling of the devices' inner workings. Indeed, it is 
impossible to upper-bound the amount of CT from first principles only or from any set of observed data $\mathcal{P}$ alone, since communicating devices can deterministically reproduce any $\mathcal{P}$,
and therefore simulate any degree of Bell
violation.

At first, this may seem an unwelcome departure from the purely DI approach (i.e. $\chi=0$). Nevertheless, our approach has the advantage over fully device-dependent approaches that only a \emph{single} parameter $\chi$ must be device-\emph{dependently} estimated to ensure that the protocol performs properly, and this same parameter is used irrespectively of the underlying physical realization. Morever, even in purely DI protocols the absence of communication cannot be deduced from the observed data alone, and to verify that there is indeed no-communication will necessarily involve putting our trust in certain general assumptions regarding the behavior of the devices, or relying on some trusted external hardware. Seen in this light, our approach is not very different from the standard (DI) one, except that instead of verifying in some trusted way that $\chi=0$, we must verify that $\chi$ is no greater than some finite value.  
Finally, we note that our approach allows as a safeguard to set $\chi$ to be greater than its expected value -- a feature that may be useful even in for purely DI protocols with (allegedly) non-communicating devices. 

Even though a maximal amount of CT $\chi$ cannot be guaranteed without some modeling of the devices, there are several ways to lower-bound $\chi$ from the observed data $\mathcal{P}$ only. If the devices were not fabricated by an
adversary and do not act maliciously, then these
lower bounds may provide good estimates of $\chi$.

A simple way to lower-bound $\chi$ in a DI manner is via the degree of violation of the
no-signaling conditions Eq. (\ref{sig}), computed from the observed data $\mathcal{P}$. 
From Eq. (\ref{sig}) it follows that $\chi\geq\delta/2N$.
Improved DI bounds are obtainable, however, reflecting
the fact that $\delta$ does not capture all of the information
contained in $\mathcal{P}$. The minimal amount of CT that is compatible with a given $\mathcal{P}$ is given by the solution of the
following optimization problem
\begin{gather}
\min_{\mathcal{Q}}\quad\chi\label{CT}\\
 \mathrm{s.t.}\quad \mathrm{Tr}\left(\rho\Pi_{ab|xy}\right)=P_{ab\mid xy},\nonumber\\
 -\chi \mathds{1}\preceq\Pi_{ab|xy}-\Pi_{a|x}\otimes\Pi_{b|y}\preceq\chi \mathds{1},\nonumber \end{gather}
which can be lower-bounded using the techniques of \cite{Navascues,Pironio 3}.
It is clear that this bound is optimal, since the optimization
runs over all possible states $\rho$ and sets of projectors $\{\Pi_{a|x}\}$,
$\{\Pi_{b|y}\}$ and $\{\Pi_{ab|xy}\}$ satisfying the constraints in
Eq. (\ref{bound 1}). That it constitutes an improvement over the bound
provided by Eq. (\ref{sig}) is seen by considering the case of post-quantum
non-signaling distributions (including those that do not violate Tsirelson's
bound \cite{Tsirelson}). Such distributions will not give rise to a non-vanishing bound via Eq.
(\ref{sig}). However, since they cannot be realized quantumly without
communication, they will give rise to a non-vanishing bound via Eq. (\ref{CT}). See Fig. 2.

\begin{figure}[t!]
\center{ \includegraphics[scale=0.64]{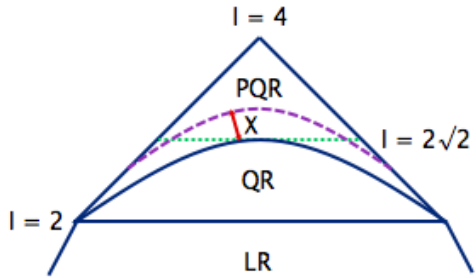}
}
\caption{Part of the no-signaling polytope. The curved solid line separates the post-quantum
region (PQR) from the quantum region (QR) -- the set of probabilities that can be realized by product measurements on quantum states. The straight dashed line represents Tsirelson’s bound. The horizontal solid
line separates QR from the classical or local region (LR). The top vertex corresponds to the PR box \cite{Popescu}.  
Although non-signaling, points in PQR cannot be realized by product measurements, but only by non-product ones. Restricting the amount of CT to  $\chi$, only points below the curved dahsed line are (quantumly) realizable.}

\end{figure}

It is possible of course that the true value of $\chi$ is not revealed
by the above lower bounds (for instance, points in QR in Fig. 2 can be reproduced either with or without CT and thus the real value of $\chi$ cannot be unambiguously determined from the observed data $\mathcal{P}$ alone). Nevertheless, one can also adopt a more device-dependent
approach to estimating $\chi$. In particular, if the lower bound provided by Eq. (\ref{CT}) equals zero,  one can vary the state
and the measurements. Such a procedure could in principle reveal the
presence of any fixed interaction Hamiltonian $H$, since it has been shown that for any such interaction there exists a strategy involving
only local operations and classical communication that reveals
the presence of the interaction as signaling \cite{Bennett}. However,
we do not know of any systematic way for finding this strategy if
$H$ is unknown, nor do we know how to relate in a systematic way
the observed signaling to $H$.

Finally, by modeling the physical systems, their interaction and
the measurement procedure, it is possible to estimate the amount of
CT. An example of this last approach are given below.

\emph{Candidates for real-life implementation} -- A system ideally suited for the implemenation of DI RG will (i) give rise to a sufficiently high Bell violation with the detection loophole kept closed, (ii) exhibit a negligible amount of CT, and (iii) allow for very high data rates.  We discuss below an experiment based on Josephson phase qubits, which meets all of these requirements. Another possibility is based on trapped ions, as discussed in Appendix B.

In the CHSH experiment of \cite{Ansmann}  two Josephson phase qubits, coupled by
a radio frequency strip resonator, are used.  The qubits are located on the
same chip, separated by $3.1\,\mathrm{mm}$, and are
entangled by successively coupling them to the strip resonator. Single
qubit rotations are effected by applying microwaves at the resonance
frequency of the corresponding qubit. Read-out is effected by letting
the excited state tunnel to an auxiliary state macroscopically distinct
from both the ground state and the excited state. All operations can
be carried out on time scales significantly shorter than $1\,\mu\mathrm{s}$.
(For a recent review of Josephson phase qubits experiments see \cite{Martinis}.)

The constant coupling
between the qubits gives rise to some CT. From the analysis of the
experimental set up performed in \cite{Ansmann} and \cite{Kofman},
it appears that the most significant contribution to the CT
occurs during the read-out: The tunneling of one qubit from the excited
state to a macroscopically distinct state sometimes forces the other
qubit to tunnel when in the ground state. This allows us to estimate the CT at $0.0030$ (see
Appendix C). The same value is also obtained by solving the second order relaxation of Eq. (\ref{CT}) using the set of observed data found in in the Supplementary Information for \cite{Ansmann}.

For the reported degree of CHSH violation $I = 2.0732$, and the above value of the CT, we find that $P_{00}^{*}\leq 0.983$. To establish robustness we note that for as a low a violation as $I=2.002$ $P_{00}^{*}\leq 0.998$.  This shows that useful randomness is extractable from this experiment.

\emph{Conclusion} -- 
The analysis of any DI protocol requires that we specify the amount of CT between the devices (irrespectively of whether it is vanishing or finite)  --  a requirement that cannot be fully verified or implemented in a DI manner. In this work we have shown that one can relax the maxims appearing previous works on DI RG, by allowing for a small amount of CT between the quantum systems. In this way we can keep most of the advantages of the DI approach and at the same time reach data rates of practical interest. Finally, we note that our approach can be generalized to other DI protocols where the devices are in the same lab, such as DI tests of genuine multi-partite entanglement \cite{Bancal}.

\begin{acknowledgments}
We acknowledge support from the European project QCS, project nbr. 255961, from the Inter-University Attraction Poles (Belgian Science Policy) project Photonics@be, from the Brussels Capital Region through a BB2B grant, and from the FRS-FNRS. The Matlab toolboxes YALMIP  \cite{YALMIP} and SeDuMi \cite{SeDuMi} were
used to obtain Fig. 2 and solve Eqs. (\ref{bound 1}) and (\ref{CT}).\end{acknowledgments}

\section*{Appendix A}

We prove here Eq. (\ref{bound 3}). 
Eq. (\ref{bound 3}) can be re-expressed as $4P^{*}_{xy}(I,\,\delta)\geq\left(4-I\right)+\left(2+8\delta\right)$.
The first term on the right-hand side is now seen to be a sum of probabilities, all of which
appear in the CHSH inequality with a minus sign: \begin{gather}
4-I  =  \sum_{a,\, b,\, x,\, y}\left[1-(-1)^{a\oplus b \oplus xy} \right]P_{ab\mid xy}\nonumber \\
  =  2\sum_{a}\left(P_{a\bar{a}\mid 00}+P_{a\bar{a}\mid 01}+P_{a\bar{a}\mid 10}+P_{aa\mid 11}\right)\,,\end{gather}
where $\bar{a} = a \oplus 1$. Hence, any probability, appearing in the CHSH inequality with a negative 
sign, is smaller or equal to $2-I/2$.

Consider now the following relation 
\begin{widetext} \begin{gather} 2+8\delta \geq  2+2\sum_{x,\, y}\left[\left(-1\right)^{y}P_{a\mid xy}+\left(-1\right)^{x}P_{b\mid xy}\right]\nonumber \\
  =  2+2\bigl(2 P_{ab\mid 00}+P_{a\bar{b}\mid 00}-P_{a\bar{b}\mid 01}+P_{a\bar{b}\mid 10}-2P_{ab\mid 11}
  -P_{a\bar{b}\mid 11}+P_{\bar{a}b\mid 00}-P_{\bar{a}b\mid 10}+P_{\bar{a}b\mid 01}-P_{\bar{a}b\mid 11}\bigr)\nonumber \\
 \geq  2+2\bigl(2 P_{ab\mid 00}-P_{a\bar{b}\mid 01}-2P_{ab\mid 11}-P_{a\bar{b}\mid 11}-P_{\bar{a}b\mid 10}-P_{\bar{a}b\mid 11}\bigr) \\
  \geq  2+2\bigl(2 P_{ab\mid 00}-P_{a\bar{b}\mid 01}-P_{ab\mid 11}-1+P_{\bar{a}\bar{b}\mid 11}-P_{\bar{a}b\mid 10}\bigr)\nonumber \\
  \geq  2\bigl(2 P_{ab\mid 00}-P_{a\bar{b}\mid 01}-P_{ab\mid 11}-P_{\bar{a}b\mid 10}\bigr)\,.\nonumber \end{gather}
  \end{widetext}
Setting $b=a$ and summing $4-I$ and $2+8\delta$, we get \begin{gather}
6-I+8\delta \geq  4P_{aa\mid 00}+2\left(P_{a\bar{a}\mid 00}+P_{\bar{a}a\mid 00}+P_{\bar{a}a\mid 01}\right.\nonumber \\
    \left. +P_{a\bar{a}\mid 10}+P_{\bar{a}\bar{a}\mid 11}\right)\,,\end{gather}
and so \begin{equation}
P_{aa\mid 00}\leq\frac{3}{2}-\frac{1}{4}I+2\delta \,.\end{equation}

 Similarly, from 
\begin{equation} 4 \delta \geq  \sum_{x,\, y}\bigl[\left(-1\right)^{x+y+1}P_{a\mid xy} + \left(-1\right)^{x}P_{b\mid xy}\bigr]\,,
\end{equation}
it follows that  
\begin{equation}
P_{aa\mid 01}\leq\frac{3}{2}-\frac{1}{4}I+2\delta \,,\end{equation} 
from  
\begin{equation} 4 \delta   \geq  \sum_{x,\, y}\bigl[\left(-1\right)^{y}P_{a\mid xy} + \left(-1\right)^{x+y+1}P_{b\mid xy}\bigr]
\end{equation}
it follows that 
 \begin{equation}
P_{aa\mid 10}\leq\frac{3}{2}-\frac{1}{4}I+2\delta \,,\end{equation}
and from 
\begin{equation} 4 \delta  \geq  \sum_{x}\Bigl[\sum_{y}\left(-1\right)^{x+y}P_{a\mid xy} + \left(-1\right)^{x+1}\bigl(P_{b\mid x0}+P_{\bar{b}\mid x1}\bigr)\Bigr]
\end{equation} it follows that 
\begin{equation}P_{a\bar{a}\mid 11}\leq\frac{3}{2}-\frac{1}{4}I+2\delta \,,\end{equation}
and so any probability appearing in the CHSH inequality with a positive sign is smaller or equal to $\frac{3}{2}-\frac{1}{4}I+2\delta $.
We therefore have that whenever $I\geq 2$ (the case $I\leq 2$ being trivial)  \begin{equation}
P^{*}_{xy}(I,\,\delta)\leq\max\Bigl\{ 2-\frac{1}{2}I,\,\frac{3}{2}-\frac{1}{4}I+2\delta  \Bigr\} =\frac{3}{2}-\frac{1}{4}I+2\delta  \,\end{equation} for all pairs of inputs $x$ and $y$.

\section*{Appendix B}

We derive here a rough estimate for the amount of CT
present in the experiments such as \cite{Monz}.
Vibrationally coupled ions in the same trap are one of the
most advanced quantum information processing systems. The system is initialized by preparing the ions in their vibrational ground state, following which they are entangled via the vibrational coupling. 
Measurements can be realized
on each ion individually: First, to choose the measurement setting, single qubit gate operations are
performed by addressing the ions individually with focused light beams.
The state of each ion is then measured in the computational basis using fluorescence. The whole process
takes $\lesssim1\,\mathrm{ms}$ and the fidelities of the gates and measurements are high ($\sim99\,\%$). (For a recent review of quantum information
processing using ion traps see \cite{Haffner}.)

In these experiments the ions are typically separated by $\sim5\,\mu\mathrm{m}$, resulting in some CT.  
It seems that the main
contribution to the CT is due to the single qubit rotations performed to select the measurement settings. Indeed, the light
beams used to address the ions have a width  of $2\,\mu\mathrm{m}$.
As a result, if the state of one ion is rotated by $\theta$ on the
Bloch sphere, the neighbouring ion will be rotated by $\varepsilon \simeq0.03\theta$
(this is the ratio of Rabi frequencies, as discussed in \cite{Haffner}). More specifically,  with no loss of generality we may assume that the ideal
entangled state shared by the parties is such that the corresponding
ideal measurements are projectors onto the states $\left|\psi_{abxy}\right\rangle =\left|a_{\varphi_{x}}\right\rangle \otimes|b_{\varphi_{y}}\rangle $,
where $\left|a_{\varphi_{x}}\right\rangle$ ($|b_{\varphi_{y}}\rangle$) denotes a state on the Bloch sphere parametrized by $\theta=\pi/2$ and $\varphi=\varphi_{x}=\left(-1\right)^{x}\pi/4$ ($\varphi=\varphi_{y}=\left(-1\right)^{y}\pi/4$). As explained above, due to the CT, the rotation of one ion by $\theta $ induces a rotation of the other by $\varepsilon \simeq 0.03 \theta $. The actual
measurements therefore consist of projectors onto $\left|\xi_{abxy}\left(\varepsilon\right)\right\rangle =\left|a_{\varphi_{x}\left(\varepsilon\right)}\right\rangle \otimes\left|b_{\varphi_{y}\left(\varepsilon\right)}\right\rangle $
with $\varphi_{x}\left(\varepsilon\right)=\left[\left(-1\right)^{x}+\left(-1\right)^{y}\varepsilon\right]\pi/4$
and $\varphi_{y}\left(\varepsilon\right)=\left[\left(-1\right)^{y}+\left(-1\right)^{x}\varepsilon\right]\pi/4$.

An upper bound on $\chi$ is given by the largest eigenvalue (up to
a sign) out of the set of eigenvalues of the 16 matrices $\left|\xi_{abxy}\left(\varepsilon\right)\right\rangle \left\langle \xi_{abxy}\left(\varepsilon\right)\right|-\left|\psi_{abxy}\right\rangle \left\langle \psi_{abxy}\right|$. Since in the CHSH experiment $\varphi_x,\,\varphi_y = \pm \pi /4$, we get that $\chi\lesssim 0.015$.

\section*{Appendix C}

We derive here the estimate for the amount of CT present in the
experiment reported in \cite{Ansmann}. Denote the ground and excited states
by $\left|0\right\rangle $ and $\left|1\right\rangle $, respectively.
Then, as explained in the main body of the text, the probability of obtaining anti-correlated
outcomes is attenuated. Specifically, let $p_{A}$ ($p_{B}$) be the
probability that the tunneling of qubit $A$ ($B$) -- i.e. obtaining
the outcome $1$ for the measurement of the state of qubit $A$ ($B$)
-- forces qubit $B$ ($A$) to tunnel when in the ground state. We
can model the presence of the CT as follows: \begin{gather}
\Pi_{00|xy}  =  \Pi_{0|x}\otimes\Pi_{0|y}\,,\nonumber\\
\mbox{\ensuremath{\Pi}}_{01|xy}  =  \left(1-p_{B}\right)\Pi_{0|x}\otimes\Pi_{1|y}\,, \\
\mbox{\ensuremath{\Pi}}_{10|xy}  =  \left(1-p_{A}\right)\Pi_{1|x}\otimes\Pi_{0|y}\,,\nonumber \\
\mbox{\ensuremath{\Pi}}_{11|xy}  =  \Pi_{1|x}\otimes\Pi_{1|y}+p_{B}\Pi_{0|x}\otimes\Pi_{1|y}+p_{A}\Pi_{1|x}\otimes\Pi_{0|y}\,,\nonumber \end{gather}
where using the notation of Appendix B $\Pi_{a|x}=\left|a_{\varphi_{x}}\right\rangle \left\langle a_{\varphi_{x}}\right|$,
etc.
To estimate the amount of CT we need to find the nearest set
of product POVMs. For simplicity, we assume them to have the form
\begin{gather}
M_{0|x}=\left(1-q_{A}\right)\Pi_{0|x}+q_{A}\Pi_{1|x}\,,\nonumber\\
 M_{0|y}=\left(1-q_{B}\right)\Pi_{0|y}+q_{B}\Pi_{1|y}\,.\end{gather}
Even if this is not the optimal choice it will still provide an upper
bound on $\chi$. Setting $p_A=0.0059$ and $p_B=0.0031$ (the values reported in the Supplementary Information for \cite{Ansmann}), the largest eigenvalue (up to a sign) out of the
set of eigenvalues of the 16 matrices $\Pi_{ab|xy}-M_{a|x}\otimes M_{b|y}$, after minimization with respect to $q_A$ and $q_B $,
equals $\simeq0.0030$ and is obtained when $q_A\simeq 0.0001$ and $q_B\simeq 0.0029$.

\end{document}